\def \SAIT #1 #2 {{\em Mem.\ Soc.\ Astron.\ It.\/} {\bf #1}, #2}
\def \MESS #1 #2 {{\em The Messenger\/} {\bf #1}, #2}
\def \ASTRNACH #1 #2 {{\em Astron. Nach.\/} {\bf #1}, #2}
\def \AAP #1 #2 {{\em Astron. Astrophys.\/} {\bf #1}, #2}
\def \AAL #1 #2 {{\em Astron. Astrophys. Lett.\/} {\bf #1}, L#2}
\def \AAR #1 #2 {{\em Astron. Astrophys. Rev.\/} {\bf #1}, #2}
\def \AAS #1 #2 {{\em Astron. Astrophys. Suppl. Ser.\/} {\bf #1}, #2}
\def \AJ #1 #2 {{\em Astron. J.\/} {\bf #1}, #2}
\def \ANNREV #1 #2 {{\em Ann. Rev. Astron. Astrophys.\/} {\bf #1}, #2}
\def \APJ #1 #2 {{\em Astrophys. J.\/} {\bf #1}, #2}
\def \APJL #1 #2 {{\em Astrophys. J. Lett.\/} {\bf #1}, L#2}
\def \APJS #1 #2 {{\em Astrophys. J. Suppl.\/} {\bf #1}, #2}
\def \APSS #1 #2 {{\em Astrophys. Space Sci.\/} {\bf #1}, #2}
\def \ASR #1 #2 {{\em Adv. Space Res.\/} {\bf #1}, #2}
\def \BAIC #1 #2 {{\em Bull. Astron. Inst. Czechosl.\/} {\bf #1}, #2}
\def \JSQRT #1 #2 {{\em J. Quant. Spectrosc. Radiat. Transfer\/} {\bf #1}, #2}
\def \MN #1 #2 {{\em Mon. Not. R. Astr. Soc.\/} {\bf #1}, #2}
\def \MEM #1 #2 {{\em Mem. R. Astr. Soc.\/} {\bf #1}, #2}
\def \PLR #1 #2 {{\em Phys. Lett. Rev.\/} {\bf #1}, #2}
\def \PASJ #1 #2 {{\em Publ. Astron. Soc. Japan\/} {\bf #1}, #2}
\def \PASP #1 #2 {{\em Publ. Astr. Soc. Pacific\/} {\bf #1}, #2}
\def \NAT #1 #2 {{\em Nature\/} {\bf #1}, #2}

\documentstyle{memsait}
\begin{opening}
\title{PRODUCTION OF GAMMA-RAYS IN BLAZARS} 
\author{W. BEDNAREK}
\institute{Department of Experimental Physics, University of \L \'od\'z,\\
ul. Pomorska 149/153, 90-236 \L \'od\'z, Poland}
\date{} 
\end{opening}

\begin{document}

\oddpagefooter{}{}{} 
\evenpagefooter{}{}{} 
\ 
\bigskip

\begin{abstract} 

Some aspects of theory of $\gamma$-ray production in blazar type active 
galaxies are discussed in context of recent observations.

\end{abstract}

\section{Introduction}

The multiwavelength spectrum of $\gamma$-ray blazars shows two prominent
broad bumps, the first one extending from radio up to optical- X-ray energy 
range, and the second one from X-rays up to at least 30 GeV and in some cases
up to $\sim 10$ TeV. Because many of $\gamma$-ray blazars show jets and 
features of superluminal motion, therefore it is usually assumed that the 
existence of jets (and accretion disks) are necessary conditions for 
$\gamma$-ray production in these sources. Three general scenarios are 
discussed in the literature. In the {\it slow jet model}, the 
$\gamma$-rays
are produced in a relativistic blob of particles moving along the jet with 
the Lorentz factor typically assumed to be of the order of $\gamma \sim 10$ 
(consistently with the radio observations of superluminal motion although on a 
parsec distance scale from the center of active galaxies). In the 
{\it fast jet model}, it is assumed that particles are moving in the inner jet 
almost rectilinearly with very high Lorentz factors $\gamma >> 10$. 
Particles are accelerated close to the inner disk or in the jet. They
are slowed down at the distance of observable radio jets as a result of strong 
energy losses or isotropization by the perpendicular component of the magnetic 
field present in the jet. The third scenario, called  the {\it collision model}, 
propose that $\gamma$-ray flares may originate when the compact objects e.g., 
clouds from the broad line region, massive stars, or debris of close supernova
explosions, collide with the jet plasma. The models applying different 
general scenarios are listed in Table~\ref{tab1}. In this paper 
we do not intend to give systematic review of all these models
but rather concentrate on  theoretical
interpretation of recent observational results obtained during 
last 2 years. The detailed discussion of earlier models has been done in 
review papers by e.g., Dermer \& Schlickeiser~(1992), Sikora~(1994), 
Dermer \& Gehrels~(1995), von Montigny et al.~(1995), Schlickeiser~(1996),
Sikora et al.~(1997).

It is usually considered that the low energy bump in the blazar's spectra 
is caused by synchrotron emission of relativistic electrons. 
However the origin of the high energy bump
is controversial. It may be formed as a result of inverse Compton scattering
(ICS) of low energy photons by relativistic electrons , or in a cascade 
process initiated by relativistic hadrons and/or leptons in the soft radiation 
and/or magnetic field. The location of the
maximum power in the high energy bump differs significantly between different
sources. This fact should reflect the basic properties of the 
mechanism of acceleration of particles and/or the propagation of produced
$\gamma$-rays in active galactic nuclei (AGN). Therefore we separate all 
$\gamma$-ray blazars in three groups called: MeV, GeV, and TeV $\gamma$-ray 
blazars.  Further we discuss these groups separately. 

Most of the models mentioned
in Tab.~1 assume that electrons play the basic role in production of 
$\gamma$-rays. Only a few of them propose that  processes responsible for
$\gamma$-ray production are of hadronic origin (Mannheim \& Biermann~1992, 
Bednarek~1993a, Coppi et al.~1993, Protheroe~1997, Dar \& Laor~1997, 
Bednarek \&
Protheroe~1997d). In these models the $\gamma$-ray emission has to be 
accompanied by neutrino emission allowing their straightforward testing
by  future large neutrino detectors.

\begin{table}[t]
 \caption{Models for $\gamma$-ray production in AGNs}
 \begin{tabular}{|l|l|l|}
   \hline
   {\bf slow jet models} & {\bf fast jet models} & 
{\bf collision models}  \\
   \hline
   Maraschi et al.~1992   & Coppi et al.~1993     & Dar \& Laor~1997  \\
   Bloom \& Marscher~1996 & Bednarek \& Kirk~1995 & Bednarek \& Protheroe \\ 
   Inoue \& Takahara~1996 & Bednarek et al.~1996a,b & ~~~1997a \\
   Mastichiadis \& Kirk~1997 & Bednarek~1997b      & Bednarek (this paper)  \\
   Mannheim \& Biermann~1992 & Bednarek~1997a      &                   \\
   Mannheim~1993          &  Bednarek \& Protheroe &                   \\ 
   Bednarek~1993a         &  ~~~1997d              &                   \\
   Henri \& Pelletier~1993 &                       &                   \\
   Marcowith et al.~1995  &                        &                   \\
   Roland \& Hermsen~1995 &                        &                   \\
   B\"ottcher \& Schlickeiser~1996 &               &                   \\
   Dermer \& Schlickeiser~1993 &                   &                   \\
   Sikora et al.~1994     &                        &                   \\
   Sikora \& Madejski~1996 &                       &                   \\
   Blandford \& Levinson~1995 &                    &                   \\
   Ghisellini \& Madau~1996 &                      &                   \\   
   Protheroe~1997                                  &                   \\
   Romanova \& Lovelace~1997 &                     &                   \\
   \hline
\end{tabular}
\label{tab1}
\end{table}

\section{GeV $\gamma$-ray blazars}

About 50 active galaxies has been detected by EGRET detector above $\sim 30$
MeV (von Montigny et al. 1995). Most of these sources are flat spectrum radio
quasars, many of them show feature of superluminal motion, and jets aligned
at small angles
towards the observer's line of sight. The $\gamma$-ray emission probably 
consists of two components from which the first one is persistent but on a 
low level and the second one is highly variable on a time scales from a part of 
a day up to weeks and months.
The $\gamma$-ray spectra of blazars are usually interpreted as a single 
power low with differential spectral index laying in the range between  
$1.4$ and $\sim 3$. There are some evidences that the spectra observed during
the flare states are flatter in comparison to the low states. The spectra show 
also the spectral break at MeV energies and there is a weak evidence of 
the spectral
cut-off above a few GeV (Pohl et al.~1997). The $\gamma$-ray luminosity of 
these sources dominate in many cases the luminosity observed in other
energy bands.

Recently a spectacular $\gamma$-ray flare has been observed from PKS 1622-297
(Mattox et al.~1997). This source shows the most rapid change of $\gamma$-ray
flux (doubling time is less than 3.8 hours) and extremely high peak luminosity 
corresponding to isotropic luminosity of $2.9\times 10^{49}$ erg s$^{-1}$. If
this emission is produced in a relativistic jet then the minimum Doppler factor 
of $\sim 8.1$ is required in order to avoid the absorption of $\gamma$-rays by 
co-spatially produced X-rays. The parameters of the flare observed from PKS 1622-297
allows us to estimate the energy density in the region of $\gamma$-ray 
production. Let us assume that the emission region is homogeneous 
and has a form of a disk with the thickness $r_\gamma$ and radius 
$\kappa r_\gamma$ (see Kirk 1997). It moves towards the observer located at the 
jet axis with the Lorentz factor $\gamma$. The energy density in the 
$\gamma$-ray emission region is

\begin{eqnarray}
\rho = {{L_\gamma t_v \Delta\Omega}\over{\pi r_\gamma^3\kappa^2}}\approx
7\times 10^{-34} L_\gamma t_v^{-2} \gamma^{-5} \kappa^{-2} {\rm ~erg~cm^{-3}},
\end{eqnarray}

\noindent
where $L_\gamma$ is measured $\gamma$-ray luminosity (for the isotropic 
case), $t_v$ is the variability time scale, $\Delta \Omega$ 
is the solid angle in which these $\gamma$-rays are emitted. We assume
$\Delta \Omega\approx 1/4\gamma^2$. The thickness of 
the emission region, $r_\gamma \approx \gamma t_v c$, is estimated based on 
the time flight arguments, and $c$ is the  velocity of light.

This energy may be transported from the central engine in the form of the 
kinetic energy of the jet plasma or in the form of the Poynting flux 
(see e.g. Romanova \& Lovelace~1997). Assuming this second possibility, and
by comparing Eq.~(1) with the required energy density of the magnetic field, 
we obtain lower limit on the magnetic field strength in the emission region
as a function of the jet 
Lorentz factor

\begin{eqnarray}
B\ge 10^{-16} L_\gamma^{1/2} t_v^{-1} \gamma^{-5/2}\kappa^{-1}~{\rm G}.
\end{eqnarray}

If we put into this formula the parameters observed during the flare in
PKS 1622-297, we get the magnetic field strength in the emission region 
as a function of its Lorentz factor which is for example:
$B > 250/\kappa$ G if $\gamma = 8$, $B > 25/\kappa$ G if $\gamma = 20$, and 
$B > 0.4/\kappa$ G if $\gamma = 100.$ 
If the value of parameter $\kappa$ is close to one, the conclusion from these 
simple considerations is the following, either

\begin{itemize}

\item the jets are fast with the Lorentz factor $\gamma \sim 100$, or 

\item the $\gamma$-rays are produced rather close to the central engine, or

\item the energy is transferred in the jet mainly in the form of kinetic energy
of the jet plasma.

\end{itemize}

In the next section we discuss briefly models proposed as possible 
explanations of $\gamma$-ray production in GeV blazars.

\subsection{Models for $\gamma$-ray production in GeV blazars}

Earlier models constructed with the purpose to explain the GeV
$\gamma$-ray emission from blazars use a soft radiation 
of external or internal origin in respect to the jet
as a target for relativistic particles. The target radiation 
of synchrotron origin, produced internally, is postulated by the 
synchrotron self-Compton (SSC) models (e.g. Maraschi et al.~1992, Bloom \&
Marscher~1996, Inoue \& Takahara~1996, Mastichiadis \& Kirk~1997), and 
by proton initiated cascade model (Mannheim \& Biermann~1992). 
The external radiation coming directly from the accretion disk is used 
as a target for particles accelerated in the jet
by Dermer \& Schlickeiser~(1993). The disk radiation which is 
quasi-isotropised by scattering on the matter distributed around the disk
is applied by Sikora et al.~(1994) and Blandford \& Levinson~(1995).
The interaction of particles with the quasi-isotropised jet radiation is proposed 
by Ghisellini \& Madau (1996). 
It is argued by Dermer \& Schlickeiser~(1994) that the $\gamma$-ray production
by electrons which  
scatter direct disk radiation is more efficient close to the accretion disk,
but at further distances the scattering of isotropized disk radiation may 
dominate. It seems that the recent observations of flares varying 
on a very short time scales
favour the production of $\gamma$-rays rather close to the disk, provided
that the jet is slow ($\gamma \sim 10$). However if the $\gamma$-rays are really
produced by scattering the quasi-isotropic radiation at farther distances from
the disk, then the assumption on a slow jet is not further relevant, and
better description of the distribution of particles in the jet is given 
by the fast jet model. 
The problem on the origin of soft target photons can become more clear
if some $\gamma$-ray blazars (of BL Lac type) would show strong X-ray flares
without accompanied TeV $\gamma$-ray flares. Such behaviour could be 
interpreted that the electrons are accelerated to sufficiently high energies
to produce TeV $\gamma$-rays but these energetic photons are absorbed by 
quasi-isotropic radiation. Different models proposed up to now
need different conditions in the source region. Therefore different radiation 
mechanisms may in fact operate in this same source but the observed spectrum 
may be dominated by only one or two of them. The above mentioned models and 
their relevance to the observations has been discussed already in the reviews 
mentioned in the Introduction. Further in this subsection we describe the  
models which were not considered in the previous reviews.

The $\gamma$-ray spectra observed from blazars may be also formed as a result
of cascades developing in the jet or close to the surface of an accretion 
disk. Such interpretation has been considered in the context of the Compton GRO
observations of blazars by Coppi et al.~(1993), Bednarek \& Kirk~(1995),
Bednarek~(1997a) and Bednarek \& Protheroe~(1997d). Coppi et al. assumed that
electrons (or protons) are injected instantaneously at certain distance, 
measured
along the jet, with very high Lorentz factors ($> 10^4$). They analyze the
cascade developing in the disk radiation applying different models
for the disk emission. The approach of Bednarek \& Kirk (1995) is different 
since in their work the importance of interaction of particles, 
with the radiation coming from the thick accretion disk, during 
acceleration in the jet is discussed. The cascade in the radiation 
and electric fields is analyzed. It is found that in certain conditions the
energy of particles gained from the electric field can be very efficiently
transferred to $\gamma$-rays.

Another cascade scenario is proposed by
Bednarek (1997a) and Bednarek \& Protheroe (1997d). Following the model
for acceleration of particles close to the accretion disk surface 
(Haswell et al.~1992), we consider the cascade developing in the disk radiation and 
magnetic fields above the disk surface. Particles can be accelerated close 
the disk surface in strong electric fields, induced in reconnection of 
magnetic fields which are additionally multiplied by the disk differential 
rotation (Haswell et al. 1992). Extremely 
high potential drops, of the order of $\sim 10^{20-21}$ eV, can appear according to 
such a model. Both types of particles (electrons and protons) can be 
accelerated in this same source in the reconnection regions of opposite 
polarity. Bednarek~(1997a) analyzes the cascades initiated by primary electrons
and defines the conditions for such cascades to occur. It is shown that
for strong disk radiation field and slightly curved reconnection region the 
acceleration of electrons is saturated by their curvature energy losses.
This process transfers very efficiently energy from electrons to extremely
high energy $\gamma$-rays. These $\gamma$-ray photons are able to initiate 
cascade in the magnetic field above the reconnection region via magnetic 
$e^\pm$ pair production and quantum synchrotron radiation. When the $e^\pm$
magnetic pair cascade becomes inefficient, the absorption of $\gamma$-rays in 
the disk radiation and the synchrotron emission of secondary $e^\pm$ pairs
determines the formation of the $\gamma$-ray spectrum escaping from 
active galaxy. The $\gamma$-ray spectra, obtained in such a model, are 
consistent with the
observations of GeV $\gamma$-ray blazars. They can extend through
the TeV $\gamma$-ray energy range if the disk temperature is relatively low.

The cascades initiated by protons accelerated in the
reconnection regions follow partly different scenario (Bednarek \& 
Protheroe~1997d). The curvature losses of protons are much smaller
than these ones suffered by  electrons, 
therefore protons are accelerated without significant losses
in a relatively small scale reconnection regions. Depending on the 
parameters of the model (magnetic field strength, temperature of the local 
disk radiation and length of the acceleration region), protons can
be injected into the disk radiation with energies corresponding to the 
maximum potential drop, or with energies limited by the condition for
saturation of the electric field by the products of the cascades initiated
by secondary $e^\pm$ pairs produced in collisions of protons with radiation.
If the radiation field in the acceleration region is strong then the 
secondary $e^\pm$ pairs can take most of the energy from the electric field
and further processes should occur as in the case considered by 
Bednarek~(1997a). The conditions for specific scenarios mentioned
above are discussed in detail in Bednarek \& Protheroe~(1997d).
The protons, injected from the acceleration region into the disk radiation 
with energies corresponding to the full available potential drop, escape
to the surrounding galaxy (if disk radiation is weak) or lose energy mainly
on pion production in collisions with the disk photons. The $e^\pm$ pairs from 
decay of pions are energetic enough to initiate magnetic $e^\pm$ pair
cascade. For the strong disk radiation the $\gamma$-rays are absorbed in the
soft radiation field and the emerging $\gamma$-ray spectra show  
cut-offs at $\sim 100$ GeV, consistently with the observations of GeV
$\gamma$-ray blazars but not with the observations of TeV $\gamma$-ray
blazars. Since the charged pions are produced in hadronic collisions,
this $\gamma$-ray emission is accompanied by the neutrino emission
with fluxes comparable to the $\gamma$-ray fluxes. The characteristic
 energies of neutrinos are of the order of $\sim 10^{17}$ eV. We have computed
the neutrino power spectrum which should accompany the strong $\gamma$-ray flare
observed from 3C 279 during June '91, and have found that the expected number of
neutrinos observed during the 10 day flare should give on 
average coincident one neutrino event in 1 km$^3$ detector with the angular
resolution of one degree. This rate is orders of magnitudes above the 
atmospheric neutrino background in such 
detector. Therefore the GeV $\gamma$-ray blazars are expected to emit detectable 
neutrino fluxes in contrary to the TeV $\gamma$-ray blazars.

The production of high energy radiation in collisions
of particles with matter seems to be less likely at present because
of the problems with finding dense enough target. However, in 
principle, particles might interact with the matter accumulated in the thick 
accretion disk as discussed by Bednarek~(1993a) or with the dense clouds from
the broad emission line region (Dar \& Laor~1997). 

The interactions of relativistic particles with matter
of the host galaxy, containing blazar type AGN, may also contribute to the
low level $\gamma$-ray emission. It seems obvious that 
because of the violent activity of the nucleus, the density of relativistic 
particles in the volume of galaxy containing active nucleus 
has to be orders of 
magnitude higher than in our own Galaxy. Provided that the relativistic 
particles
escape efficiently from the acceleration regions in the jet and a significant 
part is accumulated in the galaxy by galactic magnetic fields, we should expect
low level of $\gamma$-ray emission e.g., from nearby radio galaxies,
even if they do not show characteristic features of blazars. 
The persistent $\gamma$-ray luminosity of such radio galaxy 
 can be estimated by  using simple formula

\begin{eqnarray}
L_\gamma^{\rm RG}\approx 0.2 c \sigma_{\rm pp} \tau_{\rm b,es} n_{\rm G}
\eta L_{\rm b} \approx 1.8\times 10^{-16} \tau_{\rm b,es} n_{\rm G}
\eta L_{\rm b},
\end{eqnarray}

\noindent
where $\sigma_{\rm pp}$ is the proton-proton cross section, $c$ is the velocity 
of light, $\tau_{\rm b,es}$ is the characteristic residence time of 
relativistic protons in the host galaxy, which is taken to be equal to the 
activity phase, $\tau_{\rm b}$, of the blazar type AGN, or to the escape time 
,$\tau_{\rm es}$, of 
the protons from the host galaxy. $n_{\rm G}$ is the average density of
background matter in the host galaxy, $\eta$ is the efficiency of accumulation 
of the particles accelerated in the jet by the host galaxy, and $L_{\rm b}$
is the power in particles accelerated in the jet. For example, taking
$n_{\rm G} = 1$ cm$^{-3}$, $\tau_{\rm b,es} = 10^7$ years, and $\eta L_{\rm b}
= 10^{42}$ erg s$^{-1}$ (e.g. $\eta = 0.01$ and $L_{\rm b} = 10^{44}$ erg 
s$^{-1}$), then the persistent $\gamma$-ray luminosity of the radio galaxy is
$L_\gamma^{\rm RG}\approx 5\times 10^{40}$ erg s$^{-1}$. This is consistent
with not variable $\gamma$-ray luminosity ($\sim 8.3\times 10^{40}$ erg 
s$^{-1}$, Thompson et al.~(1995)) from the $\gamma$-ray source 2EG J1324-4317. 
This source is coincident with the closest radio galaxy Cen A 
(supposed to be misaligned blazar).

\section{TeV $\gamma$-ray blazars}

Two BL Lac type blazars, Mrk 421 and Mrk 501, has been detected at
TeV $\gamma$-ray energies by the Whipple Observatory (Punch et al.~1992,
Quinn et al.~1996). These discoveries are confirmed independently by
other experiments (Petry et al.~1996, Breslin et al.~1997, Barrau et al. 1997).
The third source of this type, 1E2344, has been recently detected with low 
significance
(Weekes~1997). The observed emission has a form of very strong flares
on time scales from days (Buckley et al.~1996, Breslin et al.~1997) up to 
a part of hours (Gaidos et al.~1996, Aharonian et al.~1997).
The spectrum of Mrk 421 is flat in the low state with  differential
spectral index $\sim 2.25\pm 0.19\pm 0.3$ between $0.4 - 4$ TeV (Mohanty et 
al.~1993). The spectrum in the high state is consistent with that one observed
in the low state and extends up to at least 8 TeV (Krennrich et al.~1997). 
The $\gamma$-ray spectra of BL Lac type blazars extend probably to much higher
energies. A small evidence (6.5$\sigma$) of the $\gamma$-ray emission 
above $\sim 50$ TeV from some nearby BL Lacs has been also 
reported.  Between them 
Mrk 421 has a significance of 3.8$\sigma$ (Meyer \& Westerhoff~1996). 
Similarly the spectrum of Mrk 501 extends up to at least 10 TeV in a high state
with the differential spectral index $2.49\pm 0.11\pm 0.25$ 
(Aharonian et al.~1997). The TeV $\gamma$-ray flares from Mrk 421 are 
simultaneous with the X-ray flares measured by the ASCA satellite 
(Buckley et al.~1996, Macomb et al.~1995). 

One of the first type of models in which the production of $\gamma$-rays in 
active galaxies has been discussed is so called  "synchrotron self-Compton 
model" (SSC). In the simplest version of this model (homogeneous model) the low 
energy part of the spectrum (from radio to X-rays) and the high energy part 
(from X-rays to $\gamma$-rays) are produced in this same homogeneous and 
isotropic region (a blob) by a single population of electrons. The electrons, 
accelerated by a shock moving in the jet, lose energy on synchrotron
process in the blob magnetic field and on inverse Compton scattering (ICS)
of these synchrotron photons to $\gamma$-ray energies (e.g. Inoue \&
Takahara~1996, Bloom \& Marscher~1996, Mastichiadis \& Kirk~1997).
Such a picture naturally explains synchronized variability at different
photon energies. Application of this model to the observations
of TeV $\gamma$-ray emission from Mrk 421 has been recently tested by
Bednarek \& Protheroe~(1997c). Based on the detected variability time scales 
of TeV $\gamma$-ray emission, observations of coincident flares in X-rays 
and TeV $\gamma$-rays, and the observed multiwavelength photon spectrum of 
Mrk 421, the constraints has been placed on the allowed
parameter space (magnetic field in the emission region and its Doppler factor)
for the homogeneous SSC model. For the 1 day flare, the magnetic field in the 
blob has to be limited to the range $\sim 0.025\div 0.15$ G, and the 
corresponding Doppler factors to the range $\sim 20.5 \div 10.7$. For the 
15 min flare these limits are following: $\sim 0.4\div 1.3$ G, and 
$\sim 37.6\div
24.$. The spectra calculated in terms of the homogeneous SSC model, for the 
marginal values of the allowed parameter space, are 
consistent with the available spectral information above $\sim 1$ TeV in 
the case of a flare varying on a 1 day time scale. However for recently 
reported very 
short 15 min flare, the calculated spectra are significantly steeper,
suggesting that the homogeneous SSC model has problems with describing 
relatively flat spectra observed up to $\sim 10$ TeV. 

The problems met by the homogeneous SSC model might be avoided if the 
assumption on spherical geometry of the blob is released. Recently Kirk~(1997) 
has proposed that the blob of relativistic electrons should have rather a form
of a thin plane with the ratio, $\kappa$, of the thickness to 
broadness much lower than one. 
For the case of 15 min flare from Mrk 421, this value has been estimated
as equal to $\kappa\approx 0.03$, based on the assumption that the variability
time scale of synchrotron emission is determined by the cooling time of
electrons (see also Takahashi 1996). However in the case when 
the period of activity of acceleration mechanism determines the flare timescale
(Bednarek \& Protheroe 1997c), than the values of magnetic field and 
aspect ratio $\kappa$, estimated by Kirk (1997), becomes a lower limit.

More complicated inhomogeneous SSC models in which the radiation at 
different energies is produced in different regions of the jet
(e.g. Ghisellini et al.~1985, Maraschi et al.~1992) are also proposed.
They are not constrained by the arguments discussed above. However they predict
the time delay between different photon energies which should be observed.

Another way of constraining the parameters of the emission region has been 
proposed by Bednarek~(1996).
In the standard model of the central engine of active galaxy, the massive black
hole is surrounded by an accretion disk whose strong thermal radiation can be 
directly observed in some objects (e.g. 3C 273). For reasonable accretion rates 
and central black hole masses, this radiation is strong enough to prevent the 
escape of high energy $\gamma$-rays if they are produced in a blob moving along
the jet (Bednarek~1993b, Becker \& Kafatos~1995, Zhang \& Cheng~1997). 
The $\gamma$-ray photons with energies above the threshold for $e^\pm$
pair production may escape from the disk radiation only from distances greater
than so called the radius $r_\gamma$ of the "$\gamma$-ray photosphere", which 
is defined by the condition that the optical depth for the absorption of 
$\gamma$-rays in the disk radiation is equal to one (e.g. Blandford \& 
Levinson~1995, Bednarek~1996).

In the case of Mrk 421 the disk emission is not directly observed. However
the parameters of the disk in this source can be estimated on statistical 
grounds, based on the observed correlation between the disk luminosity
and the radio jet luminosity at 5 GHz (Falcke et al.~1995). Such procedure,
curried out for Mrk 421 (Bednarek~1996), allows to estimate the radius of
the $\gamma$-ray photosphere for the $\gamma$-ray photons with maximum energies 
observed from Mrk 421. This information combined with the 
observations of variability time scale of radiation emitted by the blob, 
together with the assumption on the spherical symmetry of the blob, and the
information on the relativistic motion of the blob along the jet,
derived from superluminal motion in Mrk 421 ($\beta_{\rm app} = 3.8$ for the 
Hubble constant 50 km s$^{-1}$ Mpc$^{-1}$, Zhang \& Baath~(1990)), 
allows us to put  
constraints on the blob Lorentz factor as a function of the inner accretion 
disk radius (for details see Bednarek ~1996). Fig.~1 shows
the lower limits on the blob Lorentz factor in Mrk 421 for the cases of 
$\gamma$-ray flares variable on a time scale of $t_v = 1$ day during which 
photons 
are emitted with energies $E_\gamma = 50$ TeV (full curve) and for $t_v = 15$ 
min and photon energies $E_\gamma = 8$ TeV (dot-dashed curve) and $E_\gamma =
50$ TeV (dashed curve). The limits on the jet Lorentz factor, derived for 
reasonable disk inner radii, are of the order of a few tens, reaching
$\sim 100$ for the high mass black holes. Similar analysis
can be performed for other $\gamma$-ray emitting blazars with the features of
superluminal motion. For example in the case of QSO 1633+382, which is the EGRET
detected $\gamma$-ray blazar at large distance, the constraints are also more
restrictive than obtained on the base of superluminal motion alone 
(Bednarek~1996).

\begin{figure}
  \vspace{7.2cm}
\includegraphics{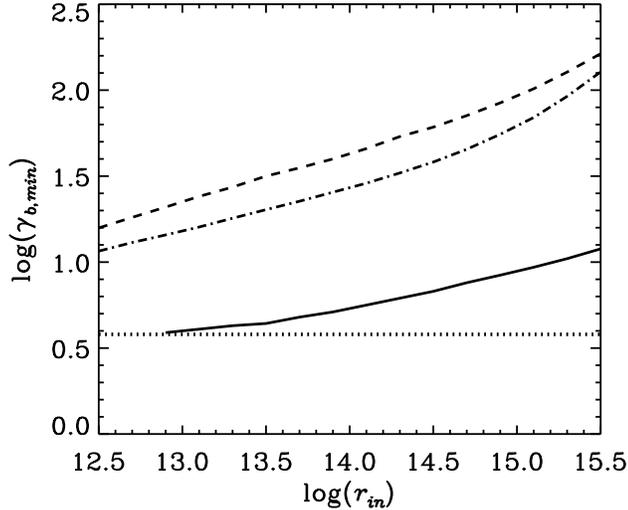}
\caption[]{Lower limits on the Lorentz factor of the blob moving in the jet of 
Mrk 421 are shown,
assuming that the $\gamma$-ray flux varies on a time scale: $t_v = 1$ day and
photons are emitted with energies $E_\gamma = 50$ TeV (full curve); $t_v = 15$
min and $E_\gamma = 8$ TeV (dot-dashed curve) and $E_\gamma = 50$ TeV (dashed 
curve). The dotted curve shows the limit obtained from observations of 
the superluminal motion alone.}
\label{fig1}
\end{figure}

\subsection{Quasi-periodic modulation of TeV $\gamma$-rays ? }

The multiwavelength campaign has been established for the observations of 
Mrk 421 during May 1995. The obtained optical, extreme UV, X-ray and 
TeV $\gamma$-ray light curves shows very interesting behaviour (see Fig.~3
in Buckley et al.~1997). At least three, and possibly four, maxima and 
corresponding minima, separated by a $\sim 3.3$ day interval, are clearly
evident in the TeV $\gamma$-ray and optical light curves 
superimposed on a broader $\sim 1$ week high activity state (see Bednarek \& 
Protheroe~1997b). The X-ray light curve shows also quasi-periodicity,
although on a day time scale, as noted by Takahashi et al. (1996) and Schubnell (1997).

This possible quasi-periodicity is not easily explained in the models which 
assume that high energy emission comes from a relativistic blob moving 
along the jet (see models mentioned in the Introduction). However most of 
the extreme UV and X-ray emission can
originate in a relatively small hot spot rotating on surface of the
inner accretion disk (Bednarek \& Protheroe~1997b). If the 
TeV emission originates in the jet above the disk then the selective absorption 
of TeV $\gamma$-rays in the hot spot radiation can induce quasi-periodic 
modulation of  TeV signal with the period corresponding to the orbital 
period of the hot spot. 
Bednarek \& Protheroe~(1997b) has computed the optical depth for TeV 
$\gamma$-rays in such scenario and put the limits on the disk radiation
below which modulation effects are not overwhelmed by the absorption in the
whole disk radiation. For Mrk 421, the possible $\sim 3.3$ day variability
could be explained if the mass of the central black hole in this source is less 
than $6\times 10^8$ M$_\odot$ (in the case of the Schwarzschild black hole) or 
$10^{10}$
M$_\odot$ (Kerr black hole). The model predicts also small time delay between
X-ray and $\gamma$-ray emission caused by the different path lengths to the 
observer from the emission regions of these radiations.

\subsection{Interactions of clouds and stars with AGN jets and $\gamma$-ray 
production.}

Dar \& Laor (1997) proposed that TeV $\gamma$-ray flares may originate in
hadronic collisions of a highly collimated relativistic proton beam 
(Lorentz factors up to $\sim 10^4$) with small, dense  clouds entering the jet. 
In this model the high level of variability of TeV $\gamma$-ray flux 
in comparison with GeV $\gamma$-ray flux is caused by stronger collimation
of TeV $\gamma$-rays in respect to GeV $\gamma$-rays if they originate
in hadronic collisions. This fact was previously noted and discussed
in astrophysical context by Bednarek et al.~(1990). However, for typical 
parameters of the clouds in AGNs (cloud temperature $\sim 10^4$ K, density
$\sim 10^{12}$ cm$^{-3}$, and radius $\sim 10^{13}$ cm), the effects of 
collisionless excitation of plasma waves by relativistic electron-proton beams
may become very efficient energy loss mechanism specially for beams with the
Lorentz factors below $\gamma_{\rm b} \sim 100$ (Rose et al.~1984). 
Therefore the GeV $\gamma$-rays can not be produced efficiently in such a 
model, since protons with  $\gamma_b \le 100$ are not able to penetrate over 
distances comparable to the typical cloud radius. Dar \& Laor (1997) has noted
that secondary $e^\pm$ pairs from decay of charged pions produce synchrotron 
X-rays and MeV - GeV $\gamma$-rays in inverse Compton process. However, in 
such a case we should expect the correlation between X-rays and MeV-GeV 
$\gamma$-rays which is not observed. The X-ray emission is correlated with the
TeV $\gamma$-ray emission in Mrk 421 (Buckley et al. 1997).

The collisions of objects with the jet in AGNs can be an attractive
mechanism of $\gamma$-ray production in another scenario proposed recently
by Bednarek \& Protheroe~(1997a). It seems obvious that many stars must collide 
with the jet plasma since the processes of star formation in the central 
regions of active galaxies are very efficient. For example, the central dense 
stellar core in the Galactic Center is $\sim 10^7 - 10^8$ stars pc$^{-3}$
(Eckart et al. 1993), and the  central total mass density has been estimated 
on $\sim 10^{9.8} M_{\odot}$ pc$^{-3}$ (Eckart \& Genzel~1996).
In M32 the central
stellar density exceeds $10^7$ M$_{\odot}$ pc$^{-3}$ (Lauer et al.~1992).
If the massive star with a strong stellar wind (e.g. Wolf-Rayet or OB type) 
enters the jet then as a result of stellar wind -- jet plasma collisions a
double shock structure should form at some distance from the star surface.
The electrons accelerated by such shocks produce X in
synchrotron process and $\gamma$-rays by scattering soft thermal 
photons coming from the 
massive star. This high energy emission is likely to be collimated along the
shock front since it is expected that the distribution of electrons 
accelerated by relativistic and oblique shock can be highly anisotropic
(Kirk \& Heavens~1989, Ostrowski~1991). The $\gamma$-ray spectra computed in 
terms of such a model can extend through the TeV energy range and the expected 
variability time scales are consistent with the observations of a two week
high level of TeV $\gamma$-ray emission observed in Mrk 421 (see Bednarek \&
Protheroe~1997a). Moreover this model suggest possible another explanation of 
quasi-periodic oscillations of TeV $\gamma$-ray flux on a time scale of days
as due to the rotation of a massive star. Rotation effect causes periodic 
variation of the magnetic fields in the region of the shocks and consequently 
influences the acceleration efficiency of electrons.

\subsection{Explosions of supernovae close to the jet}

As we have mentioned above the density of massive stars in the central
region of active galaxy ($\sim pc$) can be very high. Therefore it is 
expected that the supernova rate in central parsec is high. The
expending supernova shell can significantly compress the jet plasma if
the explosion occurs relatively close to the jet. We can estimate this
distance by comparing the pressure of the material in the supernova shell 
with the pressure of the jet plasma. The supernova pressure is
\begin{eqnarray}
P_{SN} \approx {{3M_{SN} v_{SN}^2}\over{4\pi r_{SN}^3}} \approx 
1.8\times 10^{-6} L_{50} r_1^{-3} {\rm ~erg~cm^{-3}},
\end{eqnarray}
\noindent
where $L_{50} = 10^{-50} M_{SN} v_{SN}^2/2$ is the supernova kinetic 
power in units of $10^{50}$ erg, $M_{SN}$ is the mass of the shell
and $v_{SN}$ is its velocity. $r_1$ is the radius of the shell in parsecs.
By comparing supernova shell pressure with the jet plasma pressure 
(see Eq.~1 in Bednarek \& Protheroe 1997a) we can estimate

\begin{eqnarray}
r_1 \le 0.01 L_{50}^{1/3} L_{46}^{-1/3} \theta_5^{-2/3} l_1^{-2/3} {\rm ~pc},
\end{eqnarray}

\noindent
where $L_{46}$ is the jet power in units of $10^{46}$ erg s$^{-1}$, $\theta_5$
is its opening angle in units of $5^\circ$, and $l_1$ is the distance in 
parsecs.
Note that if $r_1$ is of the order of $l_1$ then all supernovae exploding 
below $l_1$ has to interact strongly with the jet. This 
condition ($r_1\approx l_1$) defines the distance $l_{int}$ below which 
all supernovae create strong shocks in the jet,

\begin{eqnarray}
l_{int}\approx 0.06 L_{50}^{1/5} L_{46}^{-1/5} \theta_5^{-2/5} {\rm pc}.
\end{eqnarray}

\noindent
Above $l_{int}$ only a part of exploding supernovae, of the order of 
$\sim r_1^2/4l_1^2$, create strong shocks in the jet. As a result of such 
interaction a large scale ($> 10^{17}$ cm) relativistic  shock can be formed 
in the jet plasma with the magnetic field strength on the front significantly
higher than expected in isolated supernovae, because of the stronger magnetic 
field strength in the jet. In such scenario the particles can reach energies 
\begin{eqnarray}
E_{max}\approx \chi e c B r_{sh},
\end{eqnarray}
\noindent
where $\chi$ is the shock acceleration efficiency, which is of the order
of $\sim 4\times 10^{-2}$ for relativistic shocks (Protheroe~1997), $e$ is the 
elemental electric change, and $c$ is the velocity of light. 
For example if the magnetic field in the jet is $B = 1$ Gs and
$r_{sh} = 10^{17}$ cm, then particles can reach energies of the order of 
$E_{max} > 10^{18} Z$ eV, where $Z$ is the atomic number of accelerated 
nuclei. Note that $Z$ can be very high since a lot of very heavy nuclei are
expected to originate during the supernova explosion. 

If the shell front is not uniform
but contains smaller scale debris, then the number of shocks in the jet 
should be created and the particles accelerated by them may interact with the
accretion disk radiation (direct of reprocessed by a matter surrounding the disk) 
producing $\gamma$-ray photons. This picture is complementary with the 
model of $\gamma$-ray production in AGNs considered by 
Bednarek, Kirk \& Mastichiadis~(1996a,b). 

\section{MeV $\gamma$-ray blazars}

The COMPTEL telescope on the Compton GRO has detected a few blazars whose
power spectra show very strong peaks at MeV energies (Blom et al.~1995, 
Bloemen et al.~1995). It is clear that two spectral components can be 
identified in such type of sources, the MeV peak and close to the
power law spectrum extending through the EGRET energy range (Kanbach~1996).
The spectrum of the most prominent source of this type is variable with the
spectral index in the EGRET energy range changing from $1.67\pm 0.12$ to
$2.24\pm 0.36$ (Blom et al.~1996) and shows a cut-off at a few GeV 
(Kanbach~1996). 

These observations are often interpreted in terms of the $e^\pm$ pair 
annihilation
in a jet model (e.g. Marcowith et al. 1995, Roland \& Hermsen~1995, B\"ottcher
\& Schlickeiser~1996, Skibo et al.~1997). However Sikora \& Madejski~(1996) 
argue that this model can not explain the MeV excesses since
jets are optically thin and the $e^\pm$ pair annihilation is not efficient.
The annihilation model has also serious problems with explanation of the 
recent observations of 3C273 (McNaron-Brown et al.~1997). The 
spectrum of 3C 273 shows clear break at
$\sim 0.3$ MeV which is inconsistent with the blueshifted $e^\pm$ annihilation
line in a jet. Motivated by this problems, Sikora \& Madejski~(1996) proposes
that the MeV bump is produced by second population of electrons with energies
$\sim 100$ MeV present in the jet. 

Recently we have proposed another possible explanation of these intriguing
observations (Bednarek~1997b). Let us consider the general scenario in which 
electrons
are accelerated in the jet either rectilinearly in small regions along the
jet (Bednarek \& Kirk~1995, Bednarek et al.~1996a,b) or quasi-isotropically in 
a shock moving through the jet (e.g. Dermer \& Schlickeiser~1993, Sikora et 
al.~1994, Blandford \& Levinson~1995). Relativistic electrons can inverse 
Compton scatter soft radiation coming from a thin accretion disk (as considered
by Dermer \& Schlickeiser~1993 and Bednarek et al.~1996) or from a thick
accretion disk (Melia \& K\"onigl~1989 and Bednarek \& Kirk~1995). 
However if the disk geometry changes from the thin one (in the outer disk part)
to the thick one (in the inner part) then this same population of electrons
can produce strong peak at MeV energies (or below), by scattering the radiation 
of a thick disk. Close to a power law type spectrum extending above a few tens 
of MeV through the GeV energy range, originates at further distances along the 
jet as a result of  scattering the radiation of the thin disk. 
By changing only the acceleration efficiency along the jet (keeping constant 
the disk parameters) we are able to fit very different 
spectra of PKS0208-512 observed during two periods: May - June 93, and
July 91 - Jan 92. The cut-off in the spectrum observed from this source at a 
few GeV is explained by the lack of saturation of electron 
acceleration by the ICS energy losses at further distances along the jet. 
At such distances electrons are injected into the jet with the maximum possible
energies gained from acceleration regions and lose energy mainly on synchrotron
mechanism in random component of the jet magnetic field.

Two types of the soft radiation field, through which the relativistic jet 
propagates,
can be also identified with the direct disk radiation and the quasi--isotropic
radiation, reprocessed by the matter surrounding the accretion disk. These two
radiation fields has been discussed separately in earlier models.

\section{Conclusion}

The $\gamma$-ray observations of blazars have given us unique inside into the 
energetic processes occurring within central parsec of active galaxies. 
They put already very 
strong constraints on possible theoretical interpretations. Only a few general 
pictures made an effort to incorporate in a consistent model 
the variety of reported observational features like, 
very short time scale variability, different types of blazars 
(MeV, GeV, TeV), emission extending up to at least $\sim 10$ TeV.
In my opinion such a picture is given in papers which discuss the 
general model proposed initially by Dermer, Schlickeiser \& 
Mastichiadis~(1992) (and developed by Dermer \& 
Schlickeiser~1993, 1994; B\"ottcher \& Schlickeiser~1996; B\"ottcher, Mause \&
Schlickeiser~1996; Skibo, Dermer \& Schlickeiser~1997; Dermer, Sturner \& 
Schlickeiser~1997), or the model considered by Bednarek \& Kirk~(1995) (and
developed in Bednarek, Kirk \& Mastichiadis~1996a,b; Bednarek~1997b). 
Such complementary description can be found in the model proposed by Sikora, 
Begelman \& Rees~(1994) (see also, Sikora \& Madejski~1996;
Sikora, Begelman \& Madejski~1997). Some new ideas has been also investigated 
recently. They base on a less popular fast jet model
(Bednarek~1997a, Bednarek \& Protheroe~1997d), or propose that $\gamma$-rays
may originate during collisions of compact objects (clouds, massive stars, 
supernova shocks or supernova debris) with the jet plasma 
(Dar \& Laor~1997; Bednarek 
\& Protheroe~1997a; or by  Bednarek in this paper).
The simultaneous spectral information obtained recently at different energy 
ranges (although not yet completely satisfactory) suggests that the processes 
occurring there may be more complex than considered previously. It will
be not surprising if a few different 
radiation mechanisms, possibly operating at different locations are needed
in order to explain the observations. They suggest also that the inner
jets in blazars may be much faster than supposed previously based on the 
radio 
observations. Their Lorentz factors can be as high as $\sim 100$ which is 
consistent with the observations of intraday variability in some AGNs
(Begelman et al.~1994).

It is likely that part of particles, accelerated in the jet, can be trapped
inside the host blazar galaxy. This hypothesis can be checked by deep observations 
of nearby radio galaxies which are supposed to be misdirected blazars (e.g.
Cen A, Morganti et al.~ 1992). In fact the closest radio galaxy, Cen A, may be an 
example of such type of $\gamma$-ray source already detected by the EGRET 
telescope.

The real progress in understanding the $\gamma$-ray emission mechanisms
in blazars would be done with the detection (or non-detection) of the neutrino 
signals from these objects with the power comparable to their $\gamma$-ray 
power. This would decide about importance of the hadronic processes in blazars
considered by a few models (e.g. Mannheim \& Biermann 1992, Bednarek 1993, 
Coppi et al. 1993, Protheore 1997, Bednarek \& Protheroe 1997d). Some
models (Bednarek 1997, Bednarek \& Proteroe 1997d) suggest, moreover, that 
only FSRQ blazars (GeV blazars) may emit neutrinos and $\gamma$-rays, in 
contrary to BL Lac type 
blazars (TeV blazars) which may be sources of only $\gamma$-rays.

\section*{Acknowledgements}

I would like to thank the Organizing Committee of the Frascati Workshop  
for partial financial support which allowed my participation in this Workshop. 
This work is supported by the University of  \L \'od\'z grant No. 505/585.

{}

\end{document}